\documentclass[9pt,twocolumn,twoside]{pnas-new}

\articletype{Physical Sciences}
\templatetype{pnasresearcharticle}

\begin{document}

\title{A programmable stellarator-tokamak hybrid for million-scale magnetic-configuration discovery}

\author[a]{Guodong Yu}
\author[a]{Xianyi Nie}
\author[b]{Gwanggeun Seo}
\author[c]{Daxing Huang}
\author[a]{Hengqian Liu}
\author[a]{Junhao Liu}
\author[b]{Jaebeom Cho}
\author[b]{Hyun-Su Kim}
\author[a]{Jinlin Xie}
\author[a]{Ge Zhuang}
\author[c]{Fazhu Ding}
\author[b,1]{Jong-Kyu Park}
\author[a,1]{Caoxiang Zhu}

\affil[a]{CAS Key Laboratory of Frontier Physics in Controlled Nuclear Fusion, School of Nuclear Science and Technology, University of Science and Technology of China, Hefei 230027, China}
\affil[b]{Department of Nuclear Engineering, Seoul National University, Seoul 08826, Republic of Korea}
\affil[c]{Institute of Electrical Engineering, Chinese Academy of Sciences, Beijing 100190, China}

\authorcontributions{G.Y. and C.Z. designed the research; G.Y., X.N., G.S., D.H., H. L., J. L., J.C., H.-S.K., F.D., J.-K.P., and C.Z. performed the research; C.Z., J.-K.P., F.D., G.Z., and J.X. supervised the project; all authors wrote the paper.}
\authordeclaration{C.Z., G.Y., H.L., and J.L. are inventors on the patent application  202511306668.3 of the device described in the work. The other authors declare no competing interests.}
\correspondingauthor{\textsuperscript{1}To whom correspondence should be addressed. E-mail: jkpark@snu.ac.kr or caoxiangzhu@ustc.edu.cn }


\begin{abstract}
Tokamaks and stellarators are the leading magnetic-confinement concepts for fusion, but they rely on complementary design principles. Tokamaks use simple axisymmetric coils and plasma current, whereas stellarators use externally generated three-dimensional fields for steady-state operation. Here, we propose a programmable stellarator--tokamak hybrid that uses a fixed set of simple planar coils to access a broad magnetic-configuration space. The device adds 288 dipole-field coils to a tokamak-like coil set, with only six independent coil geometries required by symmetry. By programming coil currents, the same hardware generates more than 1.66 million optimized stellarator configurations spanning quasi-axisymmetry, quasi-helical symmetry, and quasi-isodynamicity, as well as tokamak-relevant three-dimensional perturbations. Representative configurations exhibit nested magnetic surfaces, low neoclassical transport, and favorable energetic-particle confinement. This approach enables rapid magnetic-configuration discovery without hardware redesign.
\end{abstract}

\doi{\url{www.pnas.org/cgi/doi/10.1073/pnas.XXXXXXXXXX}}

\maketitle
\thispagestyle{firststyle}
\ifthenelse{\boolean{shortarticle}}{\ifthenelse{\boolean{singlecolumn}}{\abscontentformatted}{\abscontent}}{}
\Firstpage

Nuclear fusion is a promising energy source for sustainable human development.
Fusion fuel, heated to the plasma state, is typically confined by magnetic fields \cite{BoozerRMP} or inertial compression \cite{NIF}.
Among various magnetic confinement fusion concepts, tokamaks and stellarators have achieved the highest fusion triple product \cite{Wurzel2025}. 
Tokamaks provide excellent confinement using axisymmetric magnetic fields produced by simple planar coils together with plasma currents, but they require sustained current drive and are vulnerable to current-driven disruptions.
Stellarators enable disruption-free steady-state operation using three-dimensional magnetic fields generated by external coils, but generic non-axisymmetric fields can suffer from large neoclassical transport.
Particle losses can be reduced by by imposing ``hidden symmetries'' in magnetic fields, like quasi-axisymmetry (QA) \cite{cfqs}, quasi-helical symmetry (QH) \cite{hsx}, and quasi-isodynamicity (QI) \cite{w7x}.
However, optimized stellarators generally require complex 3D coils that are difficult to manufacture and assemble \cite{Wechsung2022}.

Several approaches have sought to combine tokamak and stellarator advantages by adding helical coils \cite{cth}, permanent magnets \cite{Helander2020}, and ``banana coils'' \cite{Henneberg2024}.
Most prior stellarator-tokamak ``hybrids'' have been designed around conventional stellarator configurations or QA.
QI has been extensively studied as a leading candidate for future reactors.
The geometric freedom of three-dimensional fields enables broader exploration of new concepts, such as piecewise-omnigenity (pwO) \cite{Velasco2024a}and QI-pwO configurations \cite{Liu2026}.
Here, we propose a programmable stellarator-tokamak hybrid in which a fixed set of simple planar coils is current-programmed to generate a broad range of magnetic configurations.
The same hardware can access optimized stellarator fields, including QA, QH, and QI, as well as tokamak-relevant fields, including negative triangularity, edge-localized resonant magnetic perturbations (RMPs) \cite{yang2020}, and quasi-symmetric magnetic perturbations (QSMPs) \cite{parkQSMP2021}.
We have explored more than one million optimized stellarator configurations with low neoclassical transport, together with a set of advanced tokamak configurations with highly selective three-dimensional field control.
This approach separates magnetic-configuration discovery from hardware redesign, enabling rapid experimental tests of optimized stellarator and tokamak-relevant three-dimensional fields in a single device.

\Endparasplit

\section*{Results}

\subsection*{Flexible device with planar coils}
\begin{figure}[ht]
    \centering
    \includegraphics[width=\columnwidth]{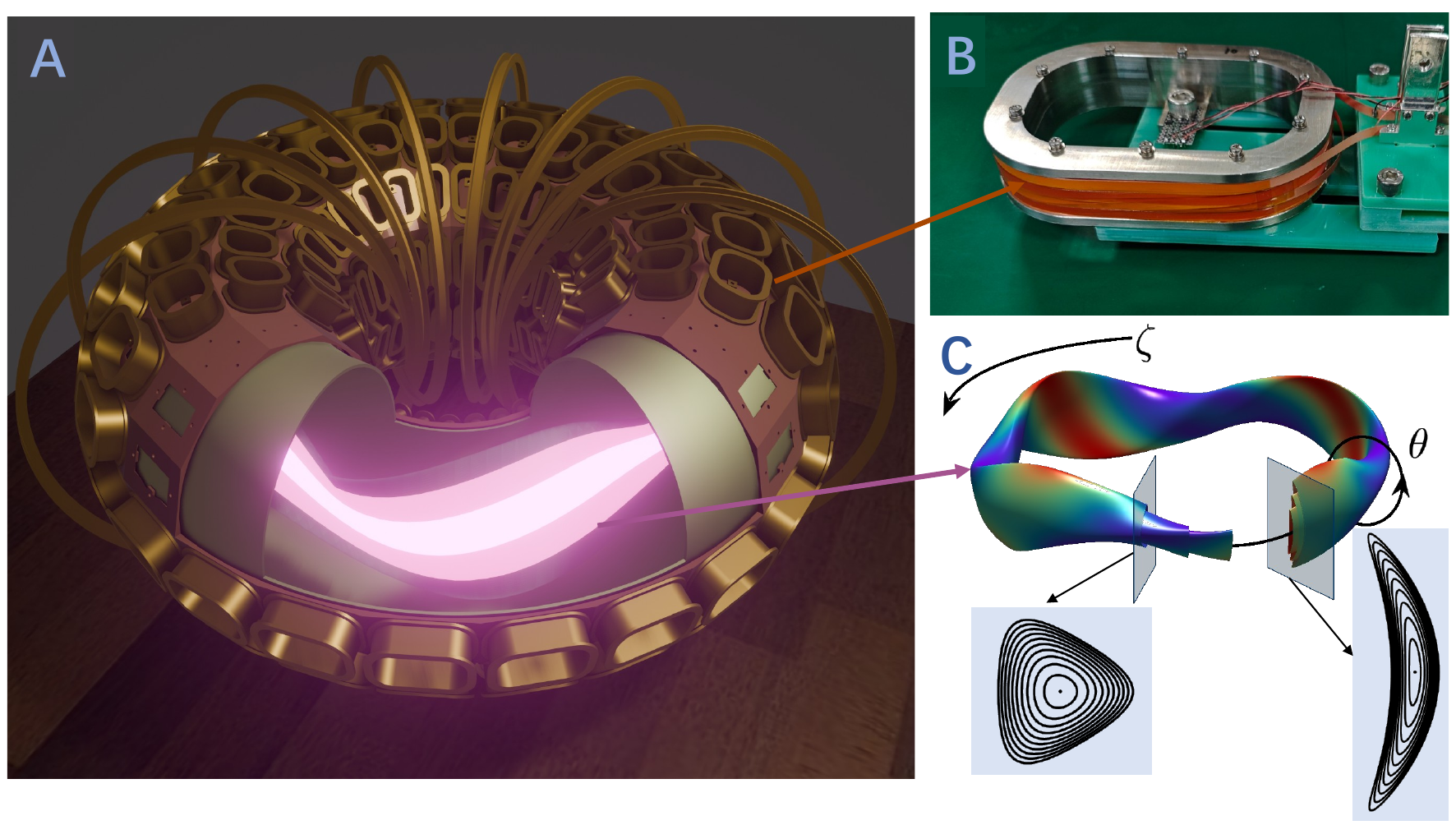}
    \caption{Programmable stellarator-tokamak hybrid. 
    (A) Schematic of the device with confined plasmas. 
    (B) Dipole-field coil fabricated with HTS tapes. 
    (C) Nested magnetic surfaces in a representative four-period QH configuration.}
    \label{fig:system_overview}
\end{figure}

Tokamaks, like ITER, typically have toroidal field (TF) coils, poloidal field (PF) coils, and a central solenoid (CS) .
The programmable stellarator-tokamak hybrid adds dipole field (DF) coils mounted on a toroidal support structure with an aspect ratio of 2.
There are 288 DF coils in total, uniformly arranged in a toroidal-by-poloidal grid of $24\times12$, as shown in Fig.~\ref{fig:system_overview}.
Owing to symmetry, only six unique coil shapes are required.
The DF coils can be fabricated from high-temperature-superconducting (HTS) tapes (Fig.~\ref{fig:system_overview}B) to carry large currents.

In stellarator mode, the PF coils and CS are not required.
The TF coils provide the background toroidal field, whereas the three-dimensional shaping fields are generated entirely by the DF coils.
By programming the DF-coil currents, the same hardware can generate many distinct stellarator configurations.
The toroidal resolution of 24 enables access of stellarator configurations with field periodicity ($n_\text{fp}$) of 2, 3, and 4.
Fig.~\ref{fig:system_overview} C shows a four-period QI configuration with nested magnetic surfaces.

\subsection*{Mega-Scale optimized stellarator configurations}
Stellarator fields must be optimized to achieve good confinement while satisfying additional physics and engineering constraints.
This is a high-dimensional, multi-objective optimization problem.
Here, the main degrees of freedom are the currents in the TF and DF coils.
A common route to reducing neoclassical transport is to impose QA, QH, or QI. 
In QA and QH configurations, the magnetic-field strength $|\mathbf{B}|$ is symmetric in the toroidal or helical direction in Boozer coordinates $(\theta_B,\zeta_B)$.
In QI configurations, $|\mathbf{B}|$ contours close poloidally, and the orbit-averaged radial drift vanishes.

\begin{figure*}[htb!]
\captionsetup{skip=0pt,font=small}
    \centering
    \includegraphics[width=\linewidth]{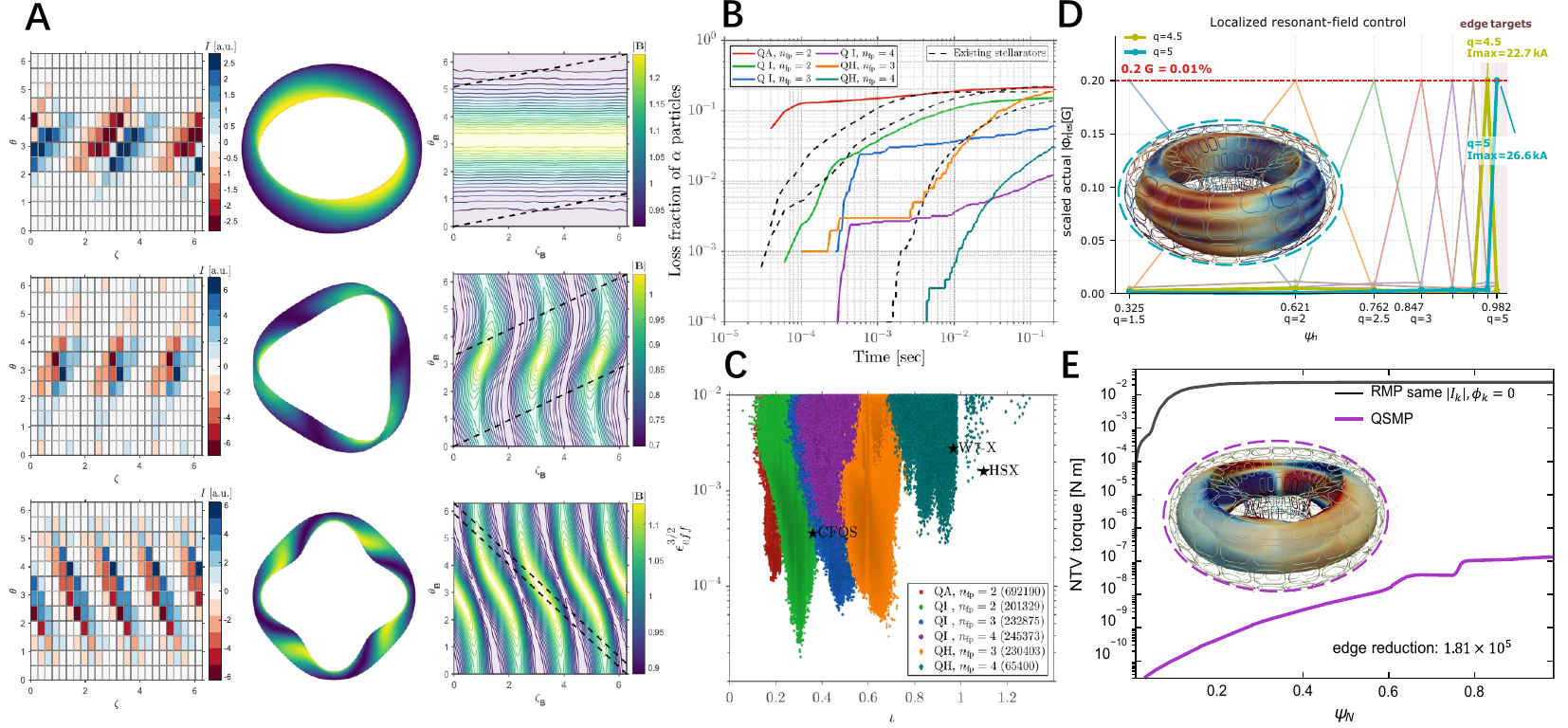}
    \caption{Optimized stellarator and tokamak configurations.
    (A) Representative optimized stellarator configurations: two-period QA (top row), three-period QI (middle row), and four-period QH (bottom row). Left: DF-coil current distributions normalized to the current in a single TF coil. Middle: three-dimensional rendering of the outermost magnetic surface, colored by magnetic-field strength. Right: $|\mathbf{B}|$ on the outermost magnetic surface in Boozer coordinates. 
    (B) Collisionless $\alpha$ particle loss fractions of six representative stellarator configurations and existing stellarators scaled to reactor size. A total of 10,000 fusion-born 3.5-MeV $\alpha$ particles are launched at normalized radial position $\psi=0.5$, and the fraction lost within 0.2 s is counted. 
    (C) Database of optimized configurations in the $\iota$ - $\epsilon_{\rm eff}^{3/2}$ parameter space. More than 1.6 million QA, QH, and QI configurations are plotted together with existing stellarators.
    (D) Individual $(m,n=2)$ resonant fields from $m=3$ to $m=10$ can be highly isolated by tailoring DF coil currents for tokamaks. 
    (E) Tokamak neoclassical toroidal viscosity is reduced by five orders of magnitude when a quasi-symmetric magnetic perturbation (QSMP) is applied.}
    \label{fig:allinone}
\end{figure*}

Fig.~\ref{fig:allinone}A presents shows representative optimized QA, QI, and QH configurations.
The DF-coil current distributions exhibit distinct patterns.
In general, inboard-side DF coils carry larger currents, whereas many DF coils carry near-zero current.
The QA configuration requires the lowest DF-coil currents, whereas the QH configuration requires the largest number of high-current DF coils.
Free-boundary equilibria are computed with the TF and DF coils.
The QA configuration has aspect ratio $A_p=6$, edge rotational transform $\iota_\text{edge}=0.19$, and boundary neoclassical transport coefficient $\epsilon_{\rm eff}^{3/2}=2.6\times10^{-4}$.
The QI configuration has $A_p=8$, $\iota_\text{edge}=0.47$, and $\epsilon_{\rm eff}^{3/2}=1.4\times10^{-4}$, whereas the QH configuration has $A_p=10$, $\iota_\text{edge}=0.92$, and $\epsilon_{\rm eff}^{3/2}=3.4\times10^{-3}$.
Collisionless $\alpha$-particle loss fractions are also evaluated after scaling several representative configurations to reactor size in Fig.~\ref{fig:allinone}B.
The four-period QI configuration exhibits the lowest $\alpha$-particle loss fraction, below 1\%, indicating good energetic-particle confinement even when discrete-coil ripple is included.

We generated a database of more than \textit{1.66 million} optimized stellarator configurations using the same fixed coil set.
The database was filtered by $\epsilon_{\rm eff}^{3/2}<10^{-2}$ and plasma aspect ratio $A_p<15$, selecting configurations with low neoclassical transport and practical compactness.
It spans QA, QH, and QI configurations with different field periodicities, and the rotational transform ranges from 0.1 to 1.3 (Fig.~\ref{fig:allinone}C).
Because all configurations are produced by programming currents in the same hardware, the database provides a direct map from coil-current space to experimentally testable magnetic configurations.

\subsection*{Advanced tokamak configurations}
The flexible device also offers unprecedented opportunities to validate non-axisymmetric (3D) field control in tokamaks. 
A critical requirement for successful 3D field control is selectivity: resonant responses should be localized to the desired rational surfaces, while non-resonant responses, especially neoclassical toroidal viscosity (NTV), should be minimized or controlled.
One example is the simultaneous minimization of resonant and non-resonant responses while preserving the desired three-dimensional shape, known as quasi-symmetric magnetic perturbation (QSMP) \cite{parkQSMP2021}.
Perturbed-equilibrium calculations show that these responses can be decoupled in the flexible device with a level of selectivity beyond the capability of present tokamak coil sets.
Fig.~\ref{fig:allinone}D shows that individual $(m,n=2)$ resonant fields can be highly isolated by tailoring the DF-coil currents and plasma shape.
This example is demonstrated in a diverted tokamak configuration with triangularity $\delta\sim0.6$ and normalized beta $\beta_N\sim1.5$, but the same controllability is achievable across a wide range of plasma parameters, including negative triangularity. 
Figure.~\ref{fig:allinone}E shows that the NTV can be reduced by five orders of magnitude when a QSMP is designed using the DF coils together with the 3D shaping.
Such decoupled responses would provide stringent tests of tokamak field-control physics and inform three-dimensional control strategies for ITER and next-step devices.


\section*{Discussion}
We have proposed a programmable stellarator-tokamak hybrid that uses a fixed set of planar coils to access a broad magnetic-configuration space.
By changing only coil currents, the device can realize more than one million optimized stellarator configurations, including QA, QH, and QI configurations, and can also operate in tokamak mode with highly selective three-dimensional field control.
The same coil set may also enable access to other toroidal concepts, such as reversed-field-pinch-like configurations, although these possibilities remain to be systematically optimized.
The central advantage of this approach is that magnetic-configuration discovery is decoupled from hardware redesign.
Once built, the device could rapidly test optimized stellarator fields, tokamak-relevant perturbations, and intermediate three-dimensional configurations in a single experimental platform.

The QH and QI configurations demonstrated here are not exact QH or QI fields.
Rather, they combine QH- or QI-like structure with pwO-like features while maintaining good confinement.
This result suggests that high-quality confinement can be achieved in a broader configuration space than that defined by exact hidden symmetry alone.
The present design is not unique.
The number of coils, coil arrangement, conductor type, and current limits can be further optimized for specific physics goals or engineering constraints.
HTS conductors provide a possible route to high-current DF coils.
The on-axis magnetic-field strength could also be increased by tilting the TF coils or placing the DF coils non-axisymmetrically \cite{Gates2025}, although such changes would reduce the configuration flexibility.
Overall, the proposed platform offers a practical path toward experimentally testing a large family of optimized stellarator and tokamak-relevant three-dimensional magnetic fields without changing the underlying hardware.


\section*{Materials and methods}
The stellarator configurations were generated using two complementary single-stage optimization workflow.
First, a quasi-single-stage optimization procedure was used to obtain high-quality seed configurations, developed with the \textsc{SIMSOPT} framework.
The plasma boundary and coil currents were iteratively adjusted so that the fixed-boundary equilibrium was consistent with the magnetic field generated by the prescribed toroidal-field (TF) and dipole-field (DF) coil set, while satisfying physics and engineering targets.
The total objective function was
\begin{align}
f_{\rm total}
= \sum_i \omega_i f_i
&= \omega_{\rm symm} f_{\rm symm}
 + \omega_{\iota} f_{\iota}
 + \omega_{\rm ap} f_{\rm ap} \nonumber \\
&+ \omega_{B_{\rm norm}} f_{B_{\rm norm}}
 + \omega_{I_{\rm max}} f_{I_{\rm max}},
\end{align}
where $\omega_i$ are prescribed weights.
Here, $f_{\rm symm}$ penalizes departure from the target QA, QH, or QI property;
$f_{\iota}$ constrains the edge rotational transform;
$f_{\rm ap}$ constrains the plasma aspect ratio;
$f_{B_{\rm norm}}$ penalizes the normal-field mismatch between the target fixed-boundary equilibrium and the magnetic field generated by the coils;
and $f_{I_{\rm max}}$ regularizes the maximum current in the DF coils.
Second, a large-scale global search was performed using the differential evolution algorithm.
The free variables were the TF and DF coil currents, and the initial populations were initialized from the seed configurations.

For the representative configurations reported in the main text, free-boundary equilibria were recomputed using \textsc{VMEC} code with the TF and DF coils.
The magnetic field was transformed to Boozer coordinates using \textsc{BOOZ\_XFORM}.
The effective ripple $\epsilon_{\rm eff}^{3/2}$ was evaluated using \textsc{NEO}.
Collisionless $\alpha$-particle losses were computed using \textsc{SIMPLE}.
Free-boundary tokamak equilibria were generated by the \textsc{TokaMaker}.
Tokamak three-dimensional field responses were evaluated using \textsc{GPEC}. 
More details are provided in the Supporting Information (SI).



\dataavail{The data supporting the findings of this study are available in Zenodo at [DOI to be added upon acceptance].}

\acknow{This work was supported by the National Natural Science Foundation of China with Grant Nos. 12405267 and 12475229, the Strategic Priority Research Program of the Chinese Academy of Sciences under Grant No. XDB0790302, the International Partnership Program of the Chinese Academy of Sciences with Grant No. 123JHZ2024248GC, and the Chinese National Fusion Project for ITER. It was also supported by the National Research Foundation of Korea (KRF) grant funded from the Ministry of Science and ICT (MSIT) of South Korea, with Grant No. RS-2024-00350293.}

\showacknow{}

\bibliography{ref}

\end{document}